\documentclass[12pt]{article}
\usepackage{times}
\usepackage{geometry}
\usepackage[utf8]{inputenc}
\usepackage{enumitem,amssymb}
\usepackage{ragged2e}
\newlist{thematic}{itemize}{8}
\usepackage{pifont}
\usepackage{graphicx}
\usepackage{multirow}
\usepackage{amsmath}
\usepackage{amssymb}
\usepackage{xspace}
\usepackage{gensymb}
\usepackage{hhline}
\usepackage{multirow}
\usepackage{booktabs}
\usepackage{hyperref}
\usepackage{cleveref}
\usepackage{xcolor}
\usepackage{array}
\usepackage{aas_macros}
\usepackage{indentfirst}

\newcommand{\neff}{\ensuremath{N_{\rm eff}}\xspace}
\newcommand{\tcm}{\ensuremath{T_{\rm cm}}\xspace}

\newcommand{\dhp}{(D/H)$_{\rm P}$}
\newcommand{\DI}{D\,\textsc{i}}
\newcommand{\HI}{H\,\textsc{i}}

\begin{document}
\rightline{LA-UR-19-22355}
{
\raggedright
\huge
Astro2020 Science White Paper \linebreak

Big Bang Nucleosynthesis and Neutrino Cosmology \linebreak
\normalsize

\noindent \textbf{Thematic Areas}

Cosmology and Fundamental Physics \linebreak

\textbf{Principal Author}

Name: Evan B. Grohs
 \linebreak
Institution: \textit{University of California, Berkeley}
 \linebreak
Email: \texttt{egrohs@berkeley.edu}
 \linebreak
Phone: +01-505-667-7495
 \linebreak

\textbf{Co-authors}

\begin{tabular}{ll}
J. Richard Bond & \textit{CITA, University of Toronto}\\
Ryan J. Cooke & \textit{Durham University}\\
George M. Fuller & \textit{University of California, San Diego}\\
Joel Meyers & \textit{Southern Methodist University}\\
Mark W. Paris & \textit{Los Alamos National Laboratory}
\end{tabular}
  \linebreak

\pagebreak

\textbf{Endorsers}

\begin{tabular}{ll}
Kevork N. Abazajian & \textit{Department of Physics and Astronomy, University of California, Irvine}\\
A. B. Balantekin & \textit{University of Wisconsin, Madison}\\
Darcy Barron & \textit{University of New Mexico}\\
Carl R. Brune & \textit{Ohio University}\\
Vincenzo Cirigliano & \textit{Los Alamos National Laboratory}\\
Alain Coc & \textit{CSNSM (CNRS IN2P3)  Orsay, France}\\
Francis-Yan Cyr-Racine & \textit{Harvard University}\\
Eleonora Di Valentino & \textit{Jodrell Bank Center for Astrophysics, University of Manchester}\\
Alexander Dolgov & \textit{INFN, University of Ferrara}\\
Olivier Dor\'e & \textit{Jet Propulsion Laboratory}\\
Marco Drewes & \textit{Universit\'e catholique de Louvain}\\
Cora Dvorkin & \textit{Department of Physics, Harvard University}\\
Alexander van Engelen & \textit{Canadian Institute for Theoretical Astrophysics}\\
Brian Fields & \textit{University of Illinois, Urbana-Champaign}\\
Raphael Flauger & \textit{University of California, San Diego}\\
Michele Fumagalli & \textit{Centre for Extragalactic Astronomy, Durham University}\\
Susan Gardner & \textit{University of Kentucky, Lexington}\\
Graciela Gelmini & \textit{University of California, Los Angeles}\\
Martina Gerbino & \textit{HEP Division, Argonne National Laboratory}\\
Steen Hannestad & \textit{Department of Physics and Astronomy, Aarhus University}\\
Wick Haxton & \textit{University of California, Berkeley}\\
Karsten Jedamzik & \textit{Laboratoire d'Univers et Particules, Universite de Montpellier II}\\
Lucas Johns & \textit{University of California, San Diego}\\
Toshitaka Kajino & \textit{National Astronomical Observatory of Japan}\\
Chad T. Kishimoto & \textit{University of San Diego}\\
Lloyd Knox & \textit{Department of Physics, University of California, Davis}\\
Arthur B. McDonald & \textit{Queen's University}\\
John O'Meara & \textit{W. M. Keck Observatory}\\
Max Pettini & \textit{Institute of Astronomy, University of Cambridge}\\
Cyril Pitrou & \textit{Institut d'Astrophysique de Paris}\\
Georg Raffelt & \textit{Max Planck Institute for Physics, Munich}\\
Martin Savage & \textit{Institute for Nuclear Theory}\\
Robert Scherrer & \textit{Vanderbilt University}\\
Shashank Shalgar & \textit{Niels Bohr International Academy}\\
Evan Skillman & \textit{University of Minnesota}\\
Friedrich-Karl Thielemann & \textit{University of Basel}\\
David Tytler & \textit{University of California, San Diego}\\
Maria Cristina Volpe & \textit{AstroParticule et Cosmologie (APC), CNRS, Universit\'e Denis Diderot}\\
Robert V. Wagoner & \textit{Stanford University}\\
Matthew J. Wilson & \textit{University of Toronto}\\
\end{tabular}
  \linebreak
  
\pagebreak

\textbf{Abstract}

There exist a range of exciting scientific opportunities for Big Bang Nucleosynthesis (BBN) in the coming decade.
BBN, a key particle astrophysics
\lq\lq tool\rq\rq\ for decades, is poised to take on new capabilities to probe
beyond standard model (BSM) physics. This development is being driven by
experimental determination of neutrino properties, new nuclear reaction
experiments, advancing supercomputing/simulation capabilities, the prospect of
high-precision next-generation cosmic microwave background (CMB) observations, and the
advent of 30m class telescopes. 

}

\pagebreak

\section{Introduction}

Big Bang Nucleosynthesis (BBN) studies in the coming decade can give us a
unique ``fossil'' record of the thermal history and evolution of the early
universe, and thereby provide new insights into beyond standard model (BSM),
neutrino, particle, and dark sector physics. BBN, coupled with the primordial
deuterium abundance determined via QSO absorption lines, gave the first
determination of the baryon content of the universe. This was later verified by
the Cosmic Microwave Background (CMB) anisotropy-determined baryon-to-photon
ratio. This represents a crowning achievement of the marriage of
nuclear and particle (neutrino) physics with astronomy. Though the BBN
enterprise is 50+ years old \cite{Cyburt:2016RMP}, it is poised to undergo a
revolution driven by high precision CMB observations~\cite{cmbs4_science_book,Ade:2018sbj,Hanany:2019lle}, the advent of 30m class
optical/near-infrared telescopes~\cite{elthires}, laboratory determination of neutrino properties and nuclear cross
sections, and by the capabilities of high performance computing. As we will
discuss below, these developments are transforming BBN into a high precision
tool for vetting BSM and dark sector physics operating in the early universe.
This tool will leverage the results of accelerator-based experiments and CMB
studies.  Moreover, it will also give constraints on light element chemical
evolution which, in turn, may give insights into the history of cosmic ray
acceleration, and so into star and galaxy formation as well.

BBN has been understood in broad-brush, and used as means to explore and
constrain particle physics possibilities, for over half a century. So what is
new and how will these observational, experimental, and computational advances
transform this enterprise? The answer lies in the anticipated precision of the
observations {\it and} the calculations. For example, upcoming CMB observations will enable high precision (order $1\%$ uncertainty)
measurements of the relativistic energy density at photon decoupling, $N_{\rm
eff}$, and the primordial helium abundance with comparable uncertainty.
Likewise, 30m class telescopes will achieve sub-percent
precision in measurements of primordial deuterium.
In turn, these high precision observations could be
leveraged into high precision constraints via advanced simulations of the weak
decoupling and nucleosynthesis processes. These calculations could also achieve
order $1\%$ uncertainty, setting up a situation where any BSM physics that
alters the time/temperature/scale factor relationship at ${\cal{O}} (1\%)$ during
the extended weak decoupling regime may be constrained by the observations.

\section{Weak Decoupling, BBN, and the Computational Challenge}

\begin{figure}[ht]
\begin{minipage}{3.9in}
\includegraphics[width=3.9in]{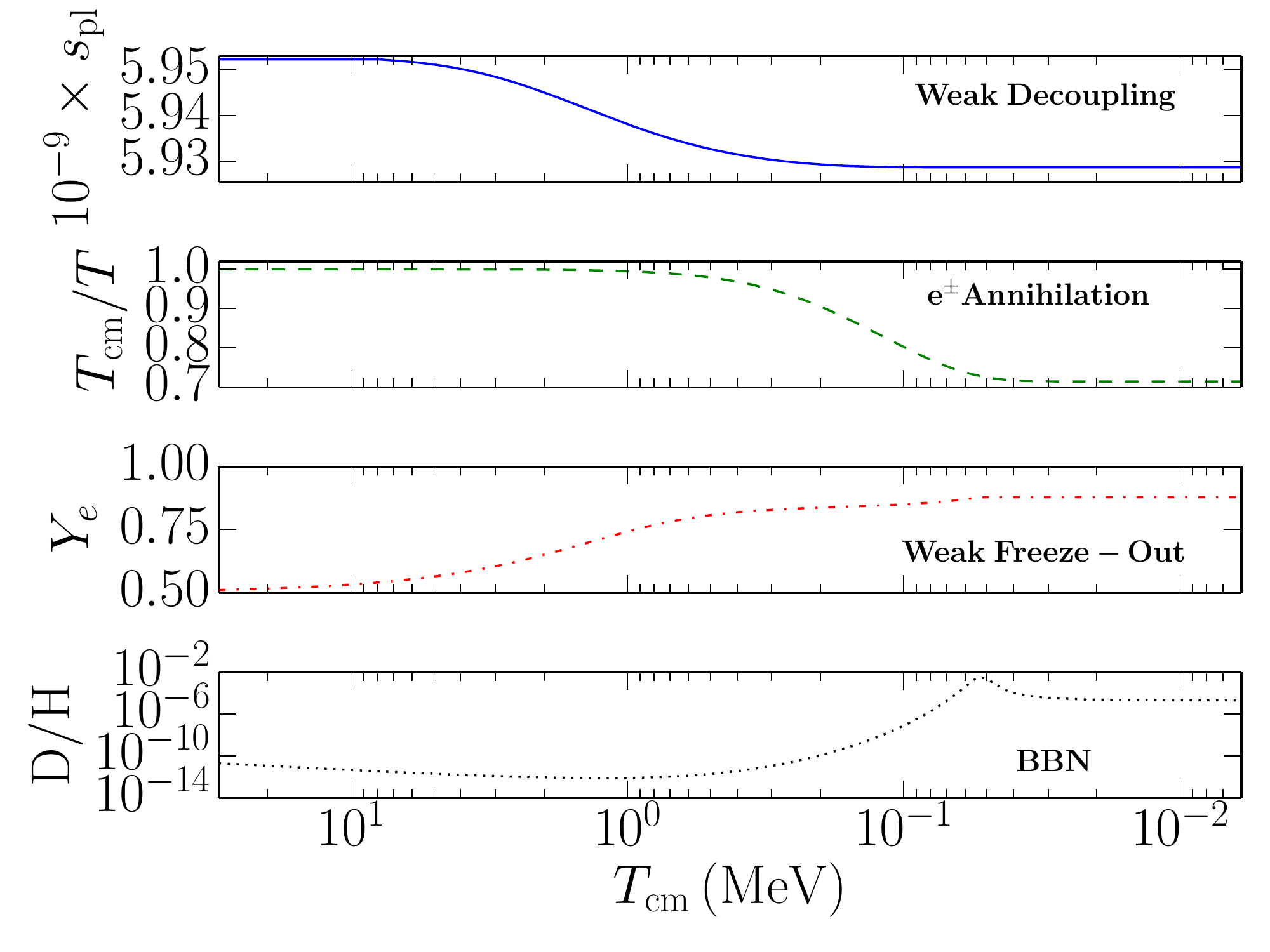}
\end{minipage}
\hfill
\begin{minipage}{2.35in}
\caption{ {\small
Entropy is transferred out of equilibrium via $\nu\-- e^\pm$ scattering from
the photon-electron-positron plasma into the decoupling neutrinos. Shown here
are the entropy per baryon in the plasma $s_{\rm pl}$ (upper panel);
temperature $T$ relative to $T_{\rm cm}$ (second from top panel), where \tcm is a proxy for scale factor; $Y_e =
1/(n/p+1)$ (third panel from top); and the deuterium abundance (bottom) as
functions of co-moving temperature $T_{\rm cm}$.
(Figure from Ref.\ \cite{Trans_BBN})
}}
\label{fig:hier}
\end{minipage}
\end{figure} 

The thermal and chemical decoupling of the neutrinos and the weak interaction,
and the closely associated freeze out of the strong and electromagnetic nuclear
reactions, together comprise a relatively lengthy process. This process plays out
over some thousands of Hubble times, $H^{-1}$, occurring roughly between
temperatures $30\,{\rm MeV} > T > 1\,{\rm keV}$.
Putative wisdom parses
this protracted process into three sequential events: (1) Weak decoupling,
wherein the rates of neutrino scattering fall below the Hubble expansion rate
$H$, implying that the exchange of energy between the neutrino and the
photon/$e^\pm$-pair plasma becomes inefficient and 
thermal equilibrium can no longer be
maintained; (2) Weak freeze out, where the charged current neutrino and lepton
capture rates fall below $H$, and nucleon isopsin, equivalently the neutron-to-proton ratio
$n/p$, drops out of equilibrium with the neutrino component; and (3) Nuclear
statistical equilibrium (NSE) freeze-out, where strong and electromagnetic
nuclear reaction rates fall below $H$ and light element abundances are frozen
in.
In fact, only recently with the advent of modern precision "kinetic" early universe large-scale parallel simulation codes has it become apparent that these three processes are not distinct, overlap for significant periods of time, and they are coupled.

%

With Standard Model physics, the evolution of the early universe through the
weak decoupling/BBN epoch is set largely by two salient features: (1) Symmetry,
in particular homogeneity and isotropy of the distribution of mass-energy; and
(2) High entropy-per-baryon.
Symmetry implies a gravitation-driven expansion which in the radiation-dominated conditions --
a consequence of high entropy  -- proceeds at rate $H
\approx {\left( 8\,\pi^3/90  \right)}^{1/2}\, g^{1/2}\,T^2/m_{\rm pl}$, where
$m_{\rm pl}$ is the Planck mass, and $g$ is the statistical weight in relativistic
particles, i.e., the photons, neutrinos, and $e^\pm$-pairs (at high temperature) of the standard model.
From the
deuterium abundance and the CMB anisotropies, we know that
the baryon-to-photon ratio is $n_{\rm b}/n_{\gamma}=\eta \approx
6.1\times{10}^{-10}$ \cite{2018arXiv180706209P}, implying an entropy-per-baryon $s \approx
5.9\times{10}^{9}$ in units of Boltzmann's constant $k_{\rm b}$.

As a homogeneous and isotropic
spacetime precludes spacelike heat flows, the co-moving entropy will be conserved when all components
and processes in the primordial plasma remain in thermal and chemical
equilibrium.
An interesting feature of the weak decoupling/BBN epoch, however,
is that they {\it do not} remain in equilibrium. The symmetry does allow for
overall homogeneous {\it timelike} entropy sources. Out of equilibrium
processes, like neutrino-electron scattering in the partially decoupled neutrino seas, provide just such an entropy source. The net result is that a small
amount of entropy is transferred from the photon-electron-positron plasma into
the decoupling neutrinos (Fig.\ \ref{fig:hier} top panel). An even smaller amount of
entropy is actually {\it generated} in this process. Note that the number of
$e^\pm$-pairs remains significant even at temperatures well below the electron
rest mass $0.511\,{\rm MeV}$ because of high entropy (Fig.\ \ref{fig:hier} second panel).

Alongside this neutrino decoupling process there is a parallel competition
between: the six (forward and reverse) isospin-changing charged current weak
interaction lepton capture/decay reactions
$\nu_e+n  \rightleftharpoons  p+e^-$,
$\bar\nu_e+p \rightleftharpoons   n+e^+$,
$n  \rightleftharpoons  p+e^-+\bar\nu_e$;
and the expansion rate $H$.
This competition, an extended freeze out process in
its own right, determines the neutron-to-proton ratio $n/p$.
The history of the baryon isospin, as expressed by the
electron fraction $Y_e = 1/(n/p +1)$, is shown as a function of co-moving
temperature in the third panel of Figure \ref{fig:hier}.

\begin{figure}[hb]
\begin{minipage}{4.3in}
\includegraphics[width=4.3in]{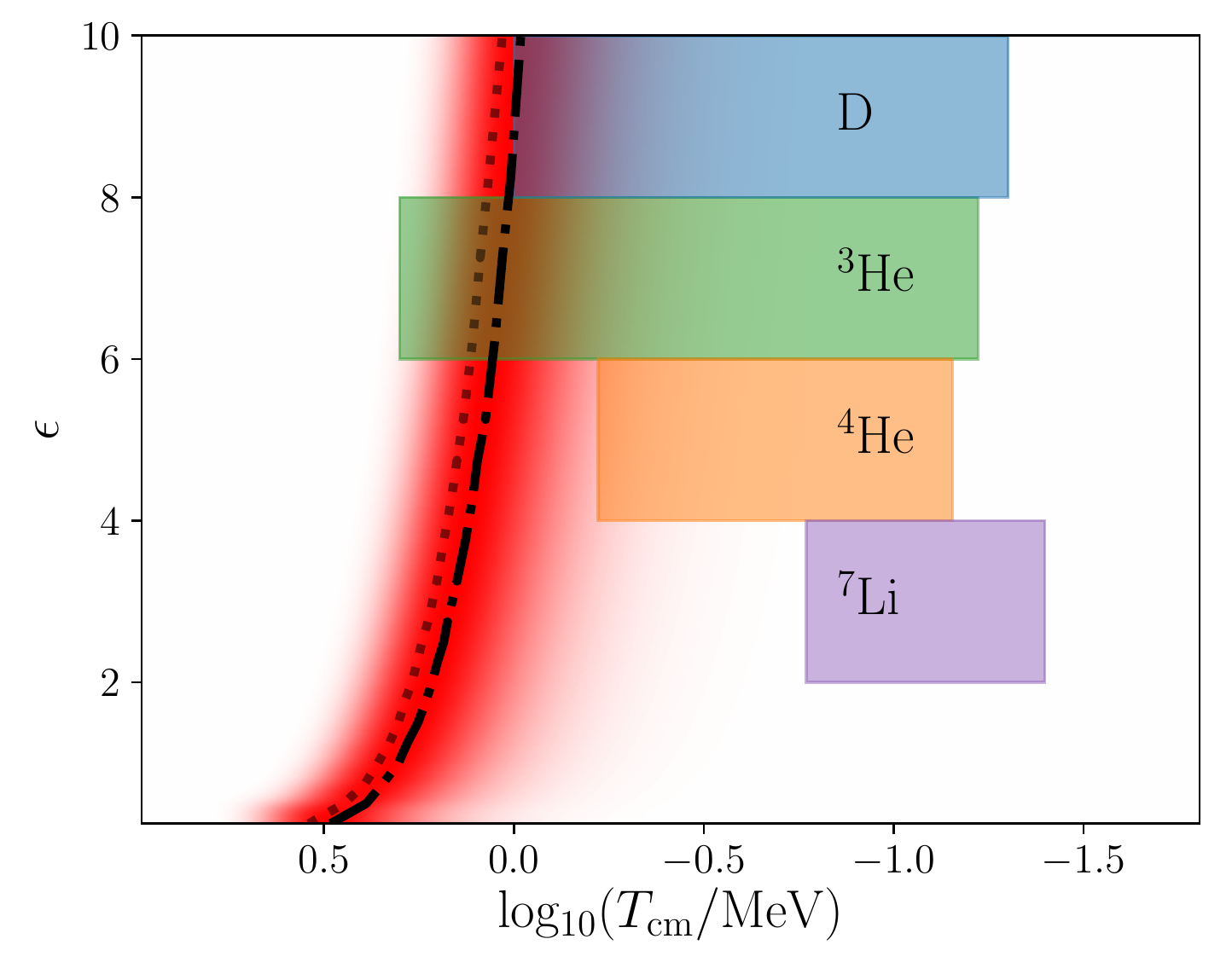}
\end{minipage}
\hfill
\begin{minipage}{2.0in}
\caption{ {\small
Simultaneous epochs of weak decoupling and BBN.  The rectangles show when in \tcm the evolving abundances of D, $^3$He, $^4$He, and $^7$Li experience rapid change either in or out of NSE.
The shaded red region shows where out-of-equilibrium electron-neutrino scattering occurs as a function of $\epsilon=E_\nu/\tcm$ for neutrino energy $E_\nu$.  The dot-dash line shows the peak kinetic transport of electron neutrinos.  The dotted line is the peak kinetic transport of either $\mu$ or $\tau$ neutrinos. c/o Matthew J.\ Wilson.
}}
\label{fig:diff_vis}
\end{minipage}
\end{figure}

References \cite{1992JETPL..56..123D,1992PhRvD..46.5378D,Dolgov:1997ne} gave the first calculations of BBN with
out-of-equilibrium neutrino spectra, whereas Ref.\ \cite{2016JCAP...07..051D}
details the most recent calculation with flavor oscillations.
Figure \ref{fig:diff_vis} shows the interplay between weak decoupling and BBN.
A new generation of early-universe simulation
calculations \cite{Trans_BBN,2016JCAP...07..051D,2002NuPhB.632..363D,neff:3.046,2015NuPhB.890..481B} take into account both the
non-equilibrium energy-transport effects, charged current weak interactions, and neutrino flavor
oscillation effects by solving the neutrino quantum kinetic equations (QKEs)
\cite{Sigl:1992fn,VFC:QKE,2015IJMPE..2441009V,Blaschke_Cirigliano_2016}. Only then is the objective
of achieving the requisite sub-percent accuracy needed for BSM signal
determination possible. This effort in precision theoretical modeling of the
early universe during BBN is then clearly essential. But it is of little use if
the nuclear reaction cross sections can not be determined with sufficient
accuracy.

Sub-percent level precision in theoretically- and experimentally-determined nuclear reaction cross
sections at low energies (a
few keV) are required to obtain sub-percent level determination of the
light-element abundances generated in BBN \cite{2018PrPNP..98...55B,2018PhRvL.121d2701D}.
Examples of nuclear reaction rate
precision determinations are given in Refs. \cite{1990ApJ...358...47K,SMK:1993bb,1998PhRvD..58f3506F,2000PhRvD..61l3505N,2015PhRvL.115m2001B,2016ApJ...831..107I,2017PhRvL.119f2002S,2017ApJ...849..134G,2018PhR...754....1P,2019PhRvC..99a4619D}. 
{\em Ab initio} \cite{2016PhRvL.116j2501M} and chiral EFT \cite{RevModPhys.81.1773} theoretical approaches are typically in the 5-10\% range
of precision for light-element capture reactions. Phenomenological R-matrix
approaches \cite{Descouvemont_2010,Paris:2014nd,Hale:2008nd}, which incorporate unitarity
constraints at the reaction amplitude level, can achieve descriptions of the
cross section precise to within a few percent of the world data with
$\chi^2$ per degree of freedom in the range from about 1.3 to 2.0.
Currently, the applied theoretical and experimental nuclear physics communities
verify and validate evaluated nuclear cross sections via a
suite of {\em integral benchmarks} \cite{Brown:2018jhj} that incorporate cross section
from large ranges of nuclides, from light-elements (H, He, Li, Be, etc.) to the
transition metals, and through the actinides. The nuclear cross section
evaluations, which are extracted from various accelerator and activation-type
experiments, are sometimes inconsistent with each other in light of the integral
benchmark constraints. For the light elements, however, the early universe
provides an excellent opportunity to constrain their interaction cross sections given the highly pure, low-A environment that obtains during BBN.

This state of affairs -- that high-energy BSM physics studies and the applied
theoretical nuclear physics evaluations of cross sections are
inextricably connected -- suggests a way forward that allows the
solution of both fundamental and applied questions.
The evaluation of the
light-element cross section data must be optimized not just to the two-body
accelerator scattering and reaction data, but also to the early universe BBN
abundances as calculated by kinetic transport codes \cite{Trans_BBN}.

\section{Terrestrial Experiments and Astronomical Observations}

What is {\it presently} known about neutrino properties is adequate, in principle, for the high fidelity weak decoupling and neutrino oscillation calculations described above.
Experiments and observations have given us the neutrino mass-squared differences and three of the parameters in the unitary transformation between the neutrino energy (mass) states and the weak interaction (flavor) states.
The measured parameters are the vacuum mixing angles, $\theta_{1 2}$, $\theta_{2 3}$, $\theta_{1 3}$, while the one (three) CP-violating phase for Dirac (Majorana) neutrinos are unknown.
Moreover, current experiments favor a normal neutrino mass hierarchy \cite{2017JCAP...06..029S} and future long baseline neutrino oscillations experiments will resolve any lingering doubt. The absolute neutrino masses remain unknown, but those quantities do not affect the BBN enterprise described here.



\begin{figure}[ht]
\begin{minipage}{4.5in}
\includegraphics[width=4.5in]{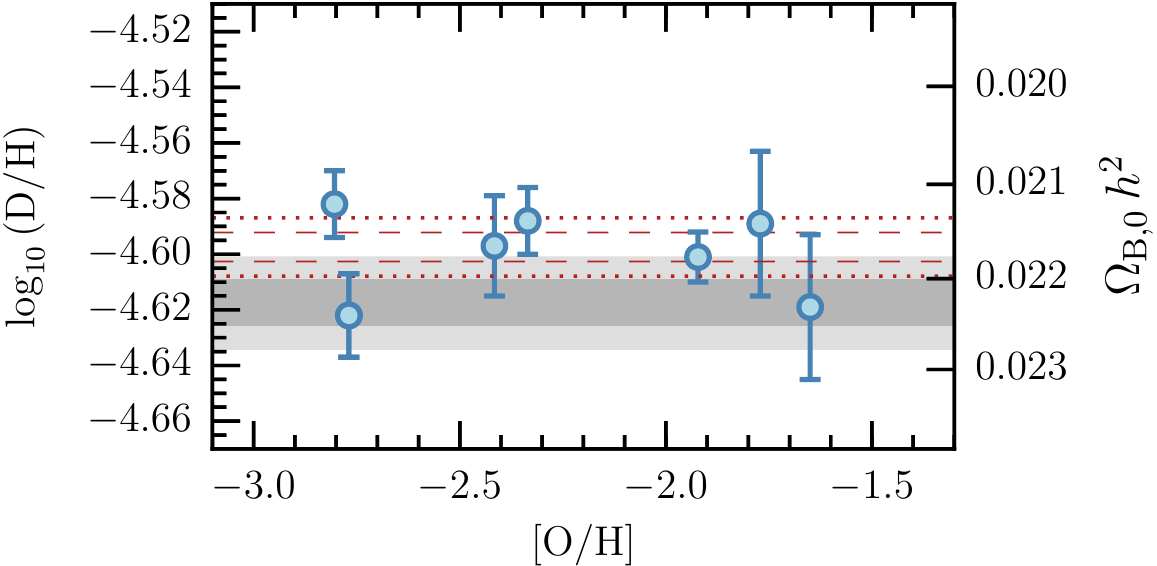}
\end{minipage}
\hfill
\begin{minipage}{1.7in}
\caption{ {\small
Measurements of D/H (blue symbols; red dashed lines show the standard error
of the weighted mean) versus [O/H] from Ref. \cite{Cooke18}. The gray band indicates the CMB value of $\Omega_{\rm
B,0}\,h^{2}$, assuming the Standard Model (dark/light bands represent 68\% and
95\% bounds).
}}
\label{fig:dh}
\end{minipage}
\end{figure}

The primordial helium and deuterium abundances are key inputs to the BBN tool for studying BSM physics.
The deuterium abundance relative to hydrogen, \dhp, is derived from gas clouds
that are seen in silhouette against an
unrelated, background quasar. The Lyman series lines (rest frame
[911--1215]\AA) of neutral deuterium and hydrogen atoms in the gas cloud absorb
the quasar's light, thereby allowing us to count the number of \DI\ and \HI\
atoms along the line-of-sight. This measurement is both accurate and precise,
however it is difficult to identify the rare, quiescent, near-pristine gas
clouds that permit the best measures (i.e. metal-poor damped Lyman-$\alpha$
systems; DLAs). The latest sample of near-pristine DLAs has allowed \dhp\ to be
measured to 1\%\ precision (see Fig.\ \ref{fig:dh} and Ref.\ \cite{Cooke18}).
However, despite two decades of research, this determination is based on only
seven systems! This meagre sample is due to: (1) The brightness of the quasars;
and (2) the accessible DLAs are limited in redshift to $2.6 \lesssim z \lesssim
3.5$ (by Earth's atmosphere, and the density of high redshift absorption
lines).

The forthcoming generation of 30\,m telescope facilities are expected to
increase the number of D/H measurements by over an order of magnitude
\cite{Cooke16}. The larger collecting area of these telescopes will allow data
to be collected for much fainter quasars, which are considerably more numerous
than the quasars that are accessible to current telescope facilities.
Furthermore, the high resolution POLLUX spectrograph onboard LUVOIR will permit
new D/H measurements of DLAs at redshifts $0.0 < z < 2.6$ towards known
$m\sim18$ quasars, in just a few hours observing time. There are two advantages
of measuring D/H at low redshift: (1) There is an increased redshift range over
which DLAs can be discovered; and (2) The Lyman-$\alpha$ forest is less dense
at low redshift, which greatly facilitates clean, reliable measurements.

The measurements of primordial helium with the lowest uncertainty [$\mathcal{O}(1\%)$] come from H {\sc ii} regions in metal-poor dwarf galaxies \cite{Aver:2013ue,2014MNRAS.445..778I}.  CMB measurements of $Y_P$ will reach a similar precision with small-scale temperature and E-mode polarization power spectra.  Although both power spectra contain degeneracies between $Y_P$ and \neff, other characteristics of cosmological spectra (in particular acoustic phase shifts in CMB \cite{Bashinsky:2003tk,Baumann:2015rya,2015PhRvL.115i1301F} and BAO \cite{Baumann:2017lmt,BAO_nu_phase_2019}) can break the degeneracy to provide meaningful measurements on $Y_P$ and \neff. 
These two parameters are independent and provide unique signatures for BSM scenarios.
Any tension between CMB and galactic-inferred $Y_P$ could motivate and inform further study of dwarf galaxy astrophysics.

\section{Future Challenges and Conclusions}

Deuterium and helium measurements agree with predictions but lithium (more specifically $^7$Li) does not at the $>3\sigma$ level \cite{Cyburt:2016RMP}.  If observations come into agreement with predictions,
then the utility of BBN for constraining BSM physics is enhanced.
If they do not, then the discrepancy could signal new physics or involve issues in stellar-evolution physics.

Solutions to the lithium problem include a class of models with a yet-to-be-determined particle.  If the sea of unknown particles decays into out-of-equilibrium standard-model particles during BBN, they can modify the nuclear photo-dissociation rates and alter the primordial abundances \cite{2004PhRvD..70f3524J}.
Such models can solve the lithium problem but create a new tension in deuterium \cite{2017IJMPE..2641002C}.
Other solutions may not require the decay of an unknown particle, but still rely on the electromagnetic plasma having distortions from thermal equilibrium \cite{2019PhRvD..99d3505K}.

Other issues in cosmology and BBN exist besides lithium.
Firstly, light sterile neutrinos as suggested by short-baseline oscillation experiments \cite{2018PhRvL.121v1801A} present a tension with current values of \neff if produced solely through oscillations with active neutrinos.  To resolve the problem, other physics would need to suppress the production, e.g., an asymmetry between neutrinos and antineutrinos or hidden interactions within the neutrino seas \cite{2018JCAP...11..049C}.  Either solution could alter the abundances away from their standard BBN values and require precise codes to characterize the deviations.
Secondly, the identity of dark matter remains elusive and its production mechanism even more nebulous.  The Hubble expansion rate during BBN is not sensitive to the energy density of dark matter, but the abundances (especially deuterium) are sensitive to the entropy in the electromagnetic sector.  If dark matter production in any way modifies the entropy history of the early universe, the abundances could provide a signature which would be absent in neutrino observables like \neff.
Lastly, dark matter and light dark fermions could be representatives of a much richer dark sector.  Dark electromagnetism \cite{2014PhRvD..90c5022F,2016JCAP...11..032B} and dark nuclear physics \cite{2014PhRvD..90k5013D,2014PhRvD..90k4506D,2018PhRvL.120v1806M} would have couplings to their standard model cousins which BBN would put tight constraints on.

%
%

There exist publicly-available codes for BBN calculations, including but not limited to PRIMAT \cite{2018PhR...754....1P}, PArthENoPE \cite{2018CoPhC.233..237C}, and AlterBBN
\cite{2018arXiv180611095A}.
These codes contain procedures to model some of the BSM physics described here.
Other codes, such as BURST \cite{Trans_BBN}, remain in development and will provide the community with additional tools to model neutrino physics in BBN.
We conclude that the BBN tool, already well used in particle astrophysics, is on the threshold of becoming a precision BSM probe.

\pagebreak

\bibliographystyle{unsrt}
\bibliography{BBN_WP_bib}

\begin{thebibliography}{10}

\bibitem{Cyburt:2016RMP}
R.~H. {Cyburt}, B.~D. {Fields}, K.~A. {Olive}, and T.-H. {Yeh}.
\newblock {Big bang nucleosynthesis: Present status}.
\newblock {\em Reviews of Modern Physics}, 88(1):015004, January 2016.

\bibitem{cmbs4_science_book}
{{\bf CMB-S4} Collaboration, J.~E. Carlstrom et al}.
\newblock {CMB-S4 Science Book, First Edition}.
\newblock {\em ArXiv e-prints}, October 2016.

\bibitem{Ade:2018sbj}
James Aguirre et~al.
\newblock {The Simons Observatory: Science goals and forecasts}.
\newblock {\em JCAP}, 1902:056, 2019.

\bibitem{Hanany:2019lle}
Shaul Hanany et~al.
\newblock {PICO: Probe of Inflation and Cosmic Origins}.
\newblock 2019.

\bibitem{elthires}
R.~Maiolino et~al.
\newblock {A Community Science Case for E-ELT HIRES}.
\newblock {\em arXiv e-prints}, October 2013.

\bibitem{Trans_BBN}
E.~{Grohs}, G.~M. {Fuller}, C.~T. {Kishimoto}, M.~W. {Paris}, and
  A.~{Vlasenko}.
\newblock {Neutrino energy transport in weak decoupling and big bang
  nucleosynthesis}.
\newblock {\em \prd}, 93(8):083522, April 2016.

\bibitem{2018arXiv180706209P}
{{\bf Planck} Collaboration}, N.~{Aghanim}, Y.~{Akrami}, M.~{Ashdown},
  J.~{Aumont}, C.~{Baccigalupi}, M.~{Ballardini}, A.~J. {Banday}, R.~B.
  {Barreiro}, N.~{Bartolo}, and et~al.
\newblock {Planck 2018 results. VI. Cosmological parameters}.
\newblock {\em ArXiv e-prints}, July 2018.

\bibitem{1992JETPL..56..123D}
A.~D. {Dolgov} and M.~{Fukugita}.
\newblock {Nonequilibrium neutrinos and primordial nucleosynthesis.}
\newblock {\em Soviet Journal of Experimental and Theoretical Physics Letters},
  56:123--126, August 1992.

\bibitem{1992PhRvD..46.5378D}
A.~D. {Dolgov} and M.~{Fukugita}.
\newblock {Nonequilibrium effect of the neutrino distribution on primordial
  helium synthesis}.
\newblock {\em \prd}, 46:5378--5382, December 1992.

\bibitem{Dolgov:1997ne}
A.~D. {Dolgov}, S.~H. {Hansen}, and D.~V. {Semikoz}.
\newblock {Non-equilibrium corrections to the spectra of massless neutrinos in
  the early universe}.
\newblock {\em Nuclear Physics B}, 503:426--444, February 1997.

\bibitem{2016JCAP...07..051D}
P.~F. {de Salas} and S.~{Pastor}.
\newblock {Relic neutrino decoupling with flavour oscillations revisited}.
\newblock {\em \jcap}, 7:051, July 2016.

\bibitem{2002NuPhB.632..363D}
A.~D. {Dolgov}, S.~H. {Hansen}, S.~{Pastor}, S.~T. {Petcov}, G.~G. {Raffelt},
  and D.~V. {Semikoz}.
\newblock {Cosmological bounds on neutrino degeneracy improved by flavor
  oscillations}.
\newblock {\em Nucl. Phys. B}, 632:363--382, June 2002.

\bibitem{neff:3.046}
G.~{Mangano}, G.~{Miele}, S.~{Pastor}, T.~{Pinto}, O.~{Pisanti}, and P.~D.
  {Serpico}.
\newblock {Relic neutrino decoupling including flavour oscillations}.
\newblock {\em Nuclear Physics B}, 729:221--234, November 2005.

\bibitem{2015NuPhB.890..481B}
J.~{Birrell}, C.~T. {Yang}, and J.~{Rafelski}.
\newblock {Relic neutrino freeze-out: Dependence on natural constants}.
\newblock {\em Nuclear Physics B}, 890:481--517, January 2015.

\bibitem{Sigl:1992fn}
G.~Sigl and G.~Raffelt.
\newblock {General kinetic description of relativistic mixed neutrinos}.
\newblock {\em Nucl. Phys.}, B406:423--451, 1993.

\bibitem{VFC:QKE}
A.~{Vlasenko}, G.~M. {Fuller}, and V.~{Cirigliano}.
\newblock {Neutrino quantum kinetics}.
\newblock {\em \prd}, 89(10):105004, May 2014.

\bibitem{2015IJMPE..2441009V}
C.~{Volpe}.
\newblock {Neutrino quantum kinetic equations}.
\newblock {\em International Journal of Modern Physics E}, 24:1541009,
  September 2015.

\bibitem{Blaschke_Cirigliano_2016}
D.~N. {Blaschke} and V.~{Cirigliano}.
\newblock {Neutrino quantum kinetic equations: The collision term}.
\newblock {\em \prd}, 94(3):033009, August 2016.

\bibitem{2018PrPNP..98...55B}
C.~{Broggini}, D.~{Bemmerer}, A.~{Caciolli}, and D.~{Trezzi}.
\newblock {LUNA: Status and prospects}.
\newblock {\em Progress in Particle and Nuclear Physics}, 98:55--84, January
  2018.

\bibitem{2018PhRvL.121d2701D}
{{\bf n TOF} Collaboration, L. Damone et al}.
\newblock {$^{7}$Be (n ,p ) $^{7}$Li Reaction and the Cosmological Lithium
  Problem: Measurement of the Cross Section in a Wide Energy Range at n\_TOF at
  CERN}.
\newblock {\em Physical Review Letters}, 121(4):042701, July 2018.

\bibitem{1990ApJ...358...47K}
L.~M. {Krauss} and P.~{Romanelli}.
\newblock {Big bang nucleosynthesis - Predictions and uncertainties}.
\newblock {\em \apj}, 358:47--59, July 1990.

\bibitem{SMK:1993bb}
M.~S. {Smith}, L.~H. {Kawano}, and R.~A. {Malaney}.
\newblock {Experimental, computational, and observational analysis of
  primordial nucleosynthesis}.
\newblock {\em \apjs}, 85:219--247, April 1993.

\bibitem{1998PhRvD..58f3506F}
G.~{Fiorentini}, E.~{Lisi}, S.~{Sarkar}, and F.~L. {Villante}.
\newblock {Quantifying uncertainties in primordial nucleosynthesis without
  Monte Carlo simulations}.
\newblock {\em \prd}, 58(6):063506, September 1998.

\bibitem{2000PhRvD..61l3505N}
K.~M. {Nollett} and S.~{Burles}.
\newblock {Estimating reaction rates and uncertainties for primordial
  nucleosynthesis}.
\newblock {\em \prd}, 61(12):123505, June 2000.

\bibitem{2015PhRvL.115m2001B}
S.~R. {Beane}, E.~{Chang}, W.~{Detmold}, K.~{Orginos}, A.~{Parre{\~n}o}, M.~J.
  {Savage}, B.~C. {Tiburzi}, and {Nplqcd Collaboration}.
\newblock {Ab initio Calculation of the n p $\rightarrow$d {$\gamma$} Radiative
  Capture Process}.
\newblock {\em Physical Review Letters}, 115(13):132001, September 2015.

\bibitem{2016ApJ...831..107I}
C.~{Iliadis}, K.~S. {Anderson}, A.~{Coc}, F.~X. {Timmes}, and S.~{Starrfield}.
\newblock {Bayesian Estimation of Thermonuclear Reaction Rates}.
\newblock {\em \apj}, 831:107, November 2016.

\bibitem{2017PhRvL.119f2002S}
M.~J. {Savage}, P.~E. {Shanahan}, B.~C. {Tiburzi}, M.~L. {Wagman}, F.~{Winter},
  S.~R. {Beane}, E.~{Chang}, Z.~{Davoudi}, W.~{Detmold}, K.~{Orginos}, and
  {Nplqcd Collaboration}.
\newblock {Proton-Proton Fusion and Tritium {$\beta$} Decay from Lattice
  Quantum Chromodynamics}.
\newblock {\em Physical Review Letters}, 119(6):062002, August 2017.

\bibitem{2017ApJ...849..134G}
{\'A}.~{G{\'o}mez I{\~n}esta}, C.~{Iliadis}, and A.~{Coc}.
\newblock {Bayesian Estimation of Thermonuclear Reaction Rates for
  Deuterium+Deuterium Reactions}.
\newblock {\em \apj}, 849:134, November 2017.

\bibitem{2018PhR...754....1P}
C.~{Pitrou}, A.~{Coc}, J.-P. {Uzan}, and E.~{Vangioni}.
\newblock {Precision big bang nucleosynthesis with improved Helium-4
  predictions}.
\newblock {\em \physrep}, 754:1--66, September 2018.

\bibitem{2019PhRvC..99a4619D}
R.~S. {de Souza}, S.~R. {Boston}, A.~{Coc}, and C.~{Iliadis}.
\newblock {Thermonuclear fusion rates for tritium + deuterium using Bayesian
  methods}.
\newblock {\em \prc}, 99(1):014619, January 2019.

\bibitem{2016PhRvL.116j2501M}
L.~E. {Marcucci}, G.~{Mangano}, A.~{Kievsky}, and M.~{Viviani}.
\newblock {Implication of the Proton-Deuteron Radiative Capture for Big Bang
  Nucleosynthesis}.
\newblock {\em Physical Review Letters}, 116(10):102501, March 2016.

\bibitem{RevModPhys.81.1773}
E.~Epelbaum, H.-W. Hammer, and Ulf-G. Mei\ss{}ner.
\newblock Modern theory of nuclear forces.
\newblock {\em Rev. Mod. Phys.}, 81:1773--1825, Dec 2009.

\bibitem{Descouvemont_2010}
P~Descouvemont and D~Baye.
\newblock {The R}-matrix theory.
\newblock {\em Reports on Progress in Physics}, 73(3):036301, feb 2010.

\bibitem{Paris:2014nd}
M.~Paris, G.~Hale, A.~Hayes-Sterbenz, and G.~Jungman.
\newblock R-matrix analysis of reactions in the 9b compound system.
\newblock {\em Nuclear Data Sheets}, 120(0):184 -- 187, 2014.

\bibitem{Hale:2008nd}
G.M. Hale.
\newblock Covariances from light-element r-matrix analyses.
\newblock {\em Nuclear Data Sheets}, 109(12):2812 -- 2816, 2008.
\newblock <ce:title>Special Issue on Workshop on Neutron Cross Section
  Covariances June 24-28, 2008, Port Jefferson, New York, USA</ce:title>.

\bibitem{Brown:2018jhj}
D.~A. Brown et~al.
\newblock {ENDF/B-VIII.0: The 8th Major Release of the Nuclear Reaction Data
  Library with CIELO-project Cross Sections, New Standards and Thermal
  Scattering Data}.
\newblock {\em Nucl. Data Sheets}, 148:1--142, 2018.

\bibitem{2017JCAP...06..029S}
Fergus {Simpson}, Raul {Jimenez}, Carlos {Pena-Garay}, and Licia {Verde}.
\newblock {Strong Bayesian evidence for the normal neutrino hierarchy}.
\newblock {\em Journal of Cosmology and Astro-Particle Physics}, 2017:029, Jun
  2017.

\bibitem{Cooke18}
R.~J. {Cooke}, M.~{Pettini}, and C.~C. {Steidel}.
\newblock {One Percent Determination of the Primordial Deuterium Abundance}.
\newblock {\em \apj}, 855:102, March 2018.

\bibitem{Cooke16}
R.~J. {Cooke}, M.~{Pettini}, K.~M. {Nollett}, and R.~{Jorgenson}.
\newblock {The Primordial Deuterium Abundance of the Most Metal-poor Damped
  Lyman-{$\alpha$} System}.
\newblock {\em \apj}, 830:148, October 2016.

\bibitem{Aver:2013ue}
E.~{Aver}, K.~A. {Olive}, R.~L. {Porter}, and E.~D. {Skillman}.
\newblock {The primordial helium abundance from updated emissivities}.
\newblock {\em \jcap}, 11:17, November 2013.

\bibitem{2014MNRAS.445..778I}
Y.~I. {Izotov}, T.~X. {Thuan}, and N.~G. {Guseva}.
\newblock {A new determination of the primordial He abundance using the He I
  {$\lambda$}10830 {\AA} emission line: cosmological implications}.
\newblock {\em \mnras}, 445:778--793, November 2014.

\bibitem{Bashinsky:2003tk}
Sergei Bashinsky and Uros Seljak.
\newblock {Neutrino perturbations in CMB anisotropy and matter clustering}.
\newblock {\em Phys. Rev.}, D69:083002, 2004.

\bibitem{Baumann:2015rya}
Daniel Baumann, Daniel Green, Joel Meyers, and Benjamin Wallisch.
\newblock {Phases of New Physics in the CMB}.
\newblock {\em JCAP}, 1601:007, 2016.

\bibitem{2015PhRvL.115i1301F}
B.~{Follin}, L.~{Knox}, M.~{Millea}, and Z.~{Pan}.
\newblock {First Detection of the Acoustic Oscillation Phase Shift Expected
  from the Cosmic Neutrino Background}.
\newblock {\em Physical Review Letters}, 115(9):091301, August 2015.

\bibitem{Baumann:2017lmt}
Daniel Baumann, Daniel Green, and Matias Zaldarriaga.
\newblock {Phases of New Physics in the BAO Spectrum}.
\newblock {\em JCAP}, 1711(11):007, 2017.

\bibitem{BAO_nu_phase_2019}
D.~{Baumann}, F.~{Beutler}, R.~{Flauger}, D.~{Green}, A.~{Slosar},
  {Vargas-Maga\~{n}a} M., B.~{Wallisch}, and C.~{Y\`{e}che}.
\newblock {First constraint on the neutrino-induced phase shift in the spectrum
  of baryon acoustic oscillations}.
\newblock {\em Nature Physics}, 15(2), February 2019.

\bibitem{2004PhRvD..70f3524J}
K.~{Jedamzik}.
\newblock {Did something decay, evaporate, or annihilate during big bang
  nucleosynthesis?}
\newblock {\em \prd}, 70(6):063524, September 2004.

\bibitem{2017IJMPE..2641002C}
A.~{Coc} and E.~{Vangioni}.
\newblock {Primordial nucleosynthesis}.
\newblock {\em International Journal of Modern Physics E}, 26:1741002, 2017.

\bibitem{2019PhRvD..99d3505K}
M.~{Kusakabe}, T.~{Kajino}, G.~J. {Mathews}, and Y.~{Luo}.
\newblock {On the relative velocity distribution for general statistics and an
  application to big-bang nucleosynthesis under Tsallis statistics}.
\newblock {\em \prd}, 99(4):043505, February 2019.

\bibitem{2018PhRvL.121v1801A}
{{\bf MiniBooNE} Collaboration, W.~C. Louis et al}.
\newblock {Significant Excess of Electronlike Events in the MiniBooNE
  Short-Baseline Neutrino Experiment}.
\newblock {\em Physical Review Letters}, 121(22):221801, November 2018.

\bibitem{2018JCAP...11..049C}
X.~{Chu}, B.~{Dasgupta}, M.~{Dentler}, J.~{Kopp}, and N.~{Saviano}.
\newblock {Sterile neutrinos with secret interactions -- cosmological discord?}
\newblock {\em \jcap}, 11:049, November 2018.

\bibitem{2014PhRvD..90c5022F}
A.~{Fradette}, M.~{Pospelov}, J.~{Pradler}, and A.~{Ritz}.
\newblock {Cosmological constraints on very dark photons}.
\newblock {\em \prd}, 90(3):035022, August 2014.

\bibitem{2016JCAP...11..032B}
J.~{Berger}, K.~{Jedamzik}, and D.~G.~E. {Walker}.
\newblock {Cosmological constraints on decoupled dark photons and dark Higgs}.
\newblock {\em \jcap}, 11:032, November 2016.

\bibitem{2014PhRvD..90k5013D}
W.~{Detmold}, M.~{McCullough}, and A.~{Pochinsky}.
\newblock {Dark nuclei. I. Cosmology and indirect detection}.
\newblock {\em \prd}, 90(11):115013, December 2014.

\bibitem{2014PhRvD..90k4506D}
W.~{Detmold}, M.~{McCullough}, and A.~{Pochinsky}.
\newblock {Dark nuclei. II. Nuclear spectroscopy in two-color QCD}.
\newblock {\em \prd}, 90(11):114506, December 2014.

\bibitem{2018PhRvL.120v1806M}
S.~D. {McDermott}.
\newblock {Is Self-Interacting Dark Matter Undergoing Dark Fusion?}
\newblock {\em Physical Review Letters}, 120(22):221806, June 2018.

\bibitem{2018CoPhC.233..237C}
R.~{Consiglio}, P.~F. {de Salas}, G.~{Mangano}, G.~{Miele}, S.~{Pastor}, and
  O.~{Pisanti}.
\newblock {PArthENoPE reloaded}.
\newblock {\em Computer Physics Communications}, 233:237--242, December 2018.

\bibitem{2018arXiv180611095A}
A.~{Arbey}, J.~{Auffinger}, K.~P. {Hickerson}, and E.~S. {Jenssen}.
\newblock {AlterBBN v2: A public code for calculating Big-Bang nucleosynthesis
  constraints in alternative cosmologies}.
\newblock {\em arXiv e-prints}, page arXiv:1806.11095, Jun 2018.

\end{thebibliography}

\end{document}